\documentclass[12pt]{article}
\usepackage{amsmath}
\usepackage{graphicx,psfrag,epsf}
\usepackage{enumerate}
\usepackage{natbib}
\usepackage{url} % not crucial - just used below for the URL 
\usepackage{color}

\newcommand{\blind}{1}

\newcommand{\calN}{\mathcal N}
\newcommand{\calF}{\mathcal F}
\newcommand{\one}{1}
\newcommand{\zero}{0}
\begin{document}

    \def\spacingset#1{\renewcommand{\baselinestretch}%
{#1}\small\normalsize} \spacingset{1}

%%%%%%%%%%%%%%%%%%%%%%%%%%%%%%%%%%%%%%%%%%%%%%%%%%%%%%%%%%%%%%%%%%%%%%%%%%%%%%

\if1\blind
{
  \title{\bf  Statistical Topology and the Random Interstellar Medium}
  \author{Robin Henderson, Irina Makarenko, Paul Bushby, \\ Andrew Fletcher, Anvar Shukurov \thanks{
    The authors gratefully acknowledge support of the Leverhulme Trust, grant RPG-2014-427}\hspace{.2cm}\\
    School of Mathematics \& Statistics, Newcastle University, UK}%\\
%    and \\
%    Author 2 \\
%    Department of ZZZ, University of WWW}
  \maketitle
} \fi

\if0\blind
{
  \bigskip
  \bigskip
  \bigskip
  \begin{center}
    {\LARGE\bf  Statistical Topology and the Random Interstellar Medium  }
\end{center}
  \medskip
} \fi

\bigskip
\begin{abstract}
Current astrophysical models of the interstellar medium assume that small scale variation and noise can be modelled as Gaussian random fields or simple transformations thereof, such as lognormal.  We use topological methods to investigate this assumption for three regions of the southern sky.  We consider  Gaussian random fields on two-dimensional lattices and investigate the expected distribution of topological structures quantified through Betti numbers.  We demonstrate that there are circumstances where differences in topology can identify differences in distributions when conventional marginal or correlation analyses may not.  We propose a non-parametric method for comparing two fields based on the counts of topological features and the geometry of the associated persistence diagrams. When we apply the  methods to the astrophysical data, we find strong evidence against a Gaussian random field model for each of the three regions of the interstellar medium that we consider. Further, we show that there are topological differences at a local scale between these different regions. 
\end{abstract}

\noindent%
{\it Keywords:}  astrophysics, Betti numbers, convex hull, filamentarity, nonparametric test, persistence diagram, random field.
\vfill

\newpage
\spacingset{1.45} % DON'T change the spacing!

%\begin{itemize}
%\item Note that figures and tables (such as Figure~\ref{fig:first} and
%Table~\ref{tab:tabone}) should appear in the paper, not at the end or
%in separate files.
%\item In the latex source, near the top of the file the command
%\verb+\newcommand{\blind}{1}+ can be used to hide the authors and
%acknowledgements, producing the required blinded version.
%\item Remember that in the blind version, you should not identify authors
%indirectly in the text.  That is, don't say ``In Smith et. al.  (2009) we
%showed that ...''.  Instead, say ``Smith et. al. (2009) showed that ...''.
%\item These points are only intended to remind you of some requirements.
%Please refer to the instructions for authors
%at \url{http://amstat.tandfonline.com/action/authorSubmission?journalCode=uasa20&page=instructions#.VFkk7fnF_0c}
%\item For more about ASA\ style, please see \url{http://journals.taylorandfrancis.com/amstat/asa-style-guide/}
%\item If you have supplementary material (e.g., software, data, technical
%proofs), identify them in the section below.  In early stages of the
%submission process, you may be unsure what to include as supplementary
%material.  Don't worry---this is something that can be worked out at later stages.
%\end{itemize}

\section{Introduction}
\label{sec:intro}

The stars of the Milky Way and other galaxies are embedded in the interstellar medium (ISM), a mixture of gas, cosmic rays and magnetic fields.   The ISM is an important and active ingredient in the Galactic system despite comprising only about $10\%$ of the total baryonic mass of the Galaxy  \citep{ferriere01}.
New stars form from cold, dense, parts of the ISM, while stellar evolution driven by energy release from thermonuclear reactions returns some of the stellar mass to the ISM via stellar winds and supernova explosions.   This injection of energy generates turbulent motions and shocks in the ISM, producing a highly heterogeneous random structure. 
Accurate knowledge of the spatial distribution of the ISM is required to understand the properties and evolution of galaxies.

\begin{figure}[t]
\begin{center}
\includegraphics[width=\textwidth]{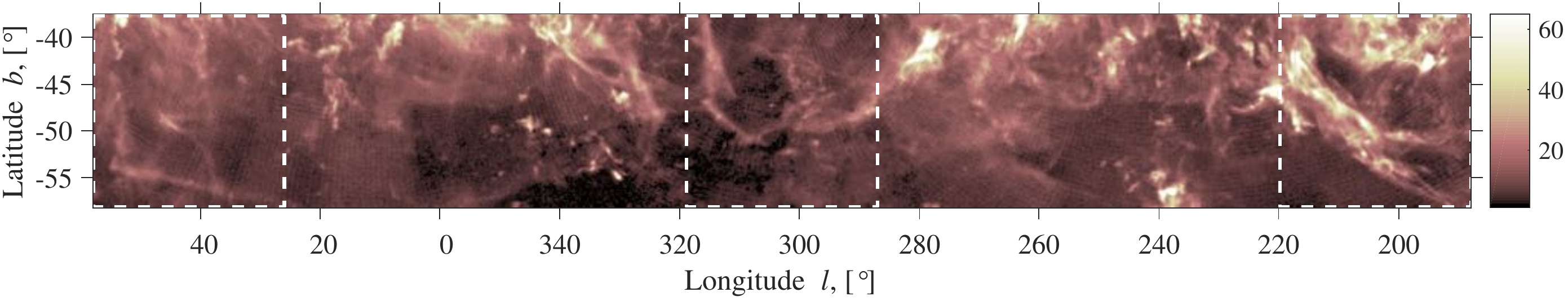}
\end{center}
\caption{\it Emission of neutral atomic hydrogen \textup{H}\,{\sc i} in a region of the Southern sky \citep{mclure09}. The coordinates are Galactic longitude $l$ and latitude $b$, centered on the Sun, where the mid-plane of the Galactic disc is at $b=0^{\circ}$ and the Galactic centre is in the direction $l=0^{\circ}$.  The brightness temperature shown is proportional to the total mass of  \textup{H}\,{\sc i} along the line of sight.  Dashed squares from left to right identify the Regions 1--3 used in Section 6.  } \label{fig:gass}
\end{figure}

One way to probe the ISM is to observe
neutral atomic hydrogen (H\,{\sc i}), as about  $90\%$  of atoms in the interstellar gas are hydrogen \citep{kalberla09}. H\,{\sc i} emits and absorbs radio waves at the frequency of $1420$\,MHz  and large data sets are now available for detailed analysis.  Figure \ref{fig:gass} shows H\,{\sc i} distribution in a section of the southern sky.  These data were obtained by the Galactic All-Sky Survey (GASS) using the Parkes $64\,{\rm m}$ radio telescope \citep{mclure09, kalberla10}.  The second and third releases of the data are available at {\tt http://www.astro.uni-bonn.de/hisurvey}. The figure shows the antenna-temperature distribution  $T(l, b)$, which is related to the gas density, as a function of position on the sky, using the coordinates of Galactic longitude, $l$, and latitude, $b$. The distance to the gas cannot be measured directly, but the Doppler-shift of the emission, dominated by the differential rotation of the Galaxy, produces a line-of-sight velocity $v$ that can in principle be used to determine the location of the gas in three dimensions. The transformation is complicated, however, and not necessary for our purposes. Instead, we obtained the two-dimensional data in Figure \ref{fig:gass}  by integrating   $T$ over  velocities from $v=20.6$ to $v=40.4\,{\rm km/s}$.

The GASS data and other surveys are rich enough to allow subtle comparisons between observations and the results of sophisticated  magneto-hydrodynamic (MHD) simulations of the ISM. The data are represented as random fields with large-scale gradients and a complex topology widely believed to be related to turbulence and outflows from the Galactic disk. At a more local level, Gaussian random fields (GRF), or simple transformations thereof, have underpinned the modelling of small-scale variations in H\,{\sc i}.  In this paper we investigate whether these models are sufficient to describe the small-scale properties of $T(l, b)$.

We consider the three regions marked by dashed lines in Figure \ref{fig:gass}, which we refer to as Regions 1--3, moving from left to right. Each consists of a $256 \times 256$ array of temperature values.  The selection of the regions was arbitrary: we did not consider any astronomical information about the locations when drawing their boundaries.  In order to concentrate on small-scale variation we removed the trend from the plots by fitting to each region a polynomial surface of order four in each of $l$ and $b$. The residuals were then marginally transformed to N(0,1).  If this transformation results in a realisation of a GRF then all information would be captured by the correlation function.  We therefore consider two questions.

\begin{enumerate}
\item[]  Q1. Are the transformed data sets consistent with stationary isotropic Gaussian random fields?
\item[]  Q2. Are there differences between the three data sets, to which the correlation function is insensitive?
\end{enumerate} 

We will address both questions using techniques  in topological data analysis, which is becoming  a popular approach 
to the analysis of random fields and more generally \citep{ ABBSW2010, AdTay2011, bubenik15, CarGun2009, Ed2014, fasy14, YA2012}. Topological invariants such as Betti numbers, the Euler characteristic, persistence diagrams and persistence barcodes, rank functions and landscapes, have been used in areas such as astrophysics \citep{Li2016arXiv}, cosmology \citep{Gay2010, Sousbie2011part1, Sousbie2011part2,  Pranav2015}, fluid dynamics \citep{Mischaikow2015arXiv, Li2016arXiv} and medicine
\citep{Davis2008, Chung2009, Chung2014arXiv}. A difference in our case compared with most previous work however is 
that we have just a single observation for each region, so that inferential techniques based on sampling and asymptotics are not appropriate.

In Section \ref{sec:sums} we describe several topological summaries that are appropriate for data on two-dimensional lattices. 
In   Section \ref{sec:grf} we study characteristics when a GRF is appropriate.
In Section \ref{sec:fields} we demonstrate that non-Gaussian random fields can sometimes be distinguished by topological features even though first and second order properties (marginal distribution and correlation function) are the same. In Section \ref{sec:tests} we propose a simple  procedure for comparing two single realisations of random fields and in Section \ref{sec:app} we describe our analysis of the GASS data.

%\begin{figure}[t]
%\begin{center}
%\includegraphics[width=2in,height=1.92in]{regionfiga.pdf}
%\includegraphics[width=2in,height=1.92in]{regionfigb.pdf}
%\includegraphics[width=2in,height=1.92in]{regionfigc.pdf}
%\end{center}
%\caption{Sample regions, after de-trending and marginal transformation. \label{fig:regions}}
%\end{figure}

\section{Topological descriptors of a random field}
\label{sec:sums}

Here we describe a number of topological measures that are suitable for analysing data that are distributed on a two-dimensional rectangular lattice.  For more general definitions and interpretations and further information see, for instance, \cite{adler10,bubenik15} or \cite{fasy14}.

\subsection{Level sets, persistent homology and persistence diagrams}

\begin{figure}[h]
\begin{center}
\includegraphics[width=4.5in,height=4.4in]{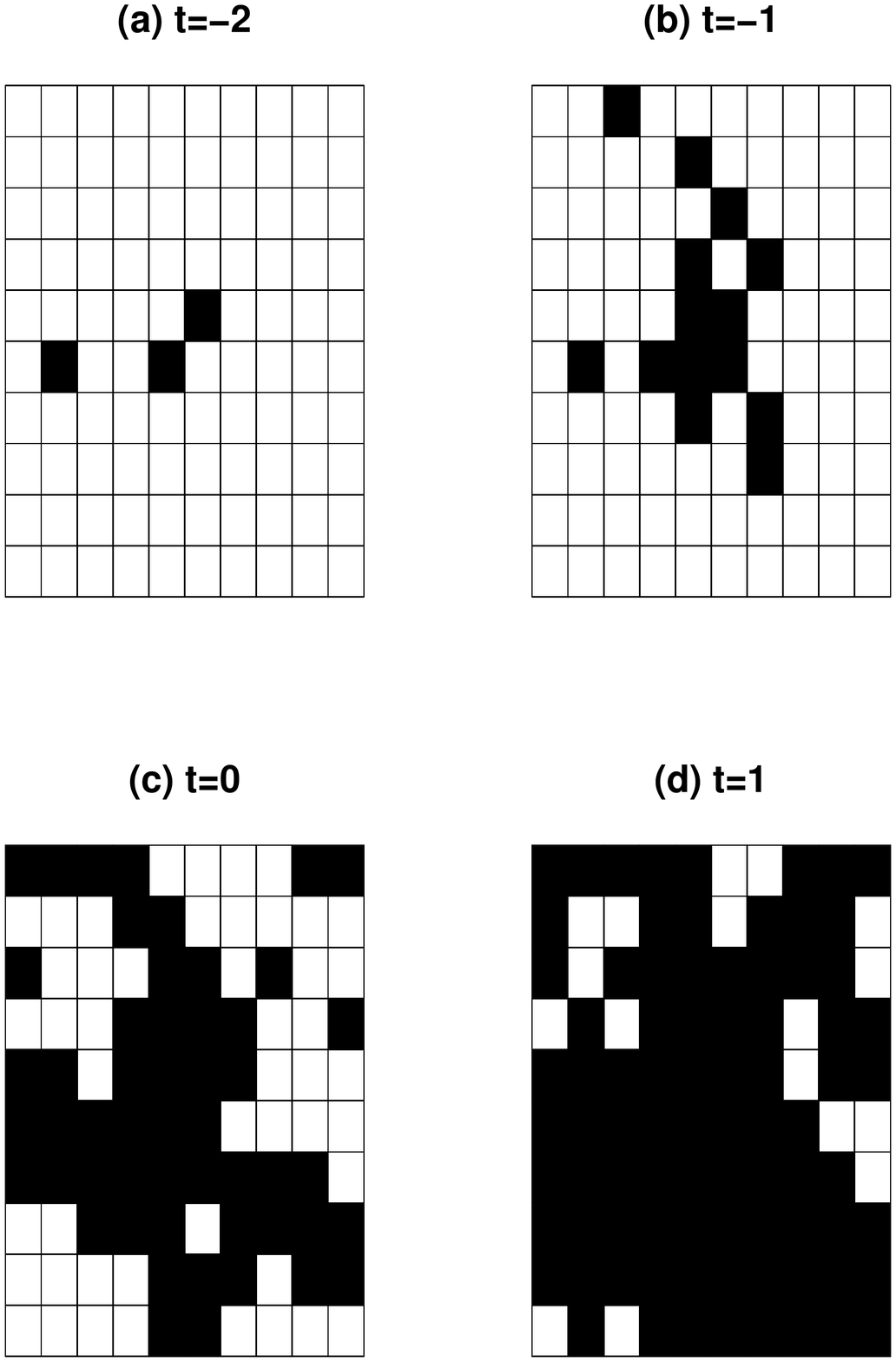}
\end{center}
\caption{{\it Four level sets (black) for a field on a $10 \times 10$ lattice.} \label{fig:grid}}
\end{figure}

Let $z(x)$ be the value of a random field at location $x$ on a two-dimensional lattice. For any real $t$, the {\it lower-level set} is defined as the locations that have field values below $t$, 
\[  \calF_t= \{x: z(x)\leq t \}.\]
Increasing $t$ from below defines a filtration which is used in {\it persistent homology} to describe the evolution of  topological structures in the field. In our context, there are two \emph{topological features} of interest, namely \emph{components} and \emph{holes}, whose counts in a level set determine  
  the {\it Betti numbers} of order zero and one, $\beta_0$ and $\beta_1$, respectively \citep{CarGun2009}.  Figure \ref{fig:grid} shows four lower-level sets for a simulated field on a $10 \times 10$ lattice and will be used to illustrate some basic concepts.  

A {\it component} is a group of one or more  pixels in a lower level set that are connected to each other, where for now we define neighbouring pixels to be connected if they have a common edge, and non-neighbouring pixels
to be connected if there is a path of connected neighbours between them. The Betti number of order zero, $\beta_0$,  for $\calF_t$ is  the number of components in the level set.  In Figure \ref{fig:grid} for instance, in panel (a) at $t=-2$ we have $\beta_0=3$, as we have assumed that pixels that share only a vertex are not connected.  By $t=-1$ and panel (b) we have $\beta_0=7$, and then $\beta_0=5$ and $\beta_0=1$ in panels (c) and (d) respectively. The emergence of a new component is described as a {\it birth} and the merger of two components is interpreted as the continuation of the component with the earlier birth time and the {\it death} of the other.  In a two-dimensional field each local minimum is associated with the birth of a component.

A {\it hole} is a group of one or more pixels that are not in the level set, are connected to each other but are isolated from other pixels that are also outside the level set.  The Betti number of order one, $\beta_1$, for $\calF_t$ is  the number of holes in the level set.  Thus in  panel (d) of Figure  \ref{fig:grid} we have $\beta_1=9$ as we ignore common vertices.  In panel  (c) we have $\beta_1$=5,   in (b) $\beta_1=2$ and in (a) we have $\beta_1=1$.  New holes are created when existing ones split, and again we can define their birth and death levels $t$. The death of a hole is associated with a local maximum. Holes could alternatively be defined by symmetry as components in a similarly defined {\it upper} level set, with the filtration now running from high to low $t$.  We will always refer to lower level sets so as to  avoid confusion.

\begin{figure}
\begin{center}
\includegraphics[width=6in,height=3.5in]{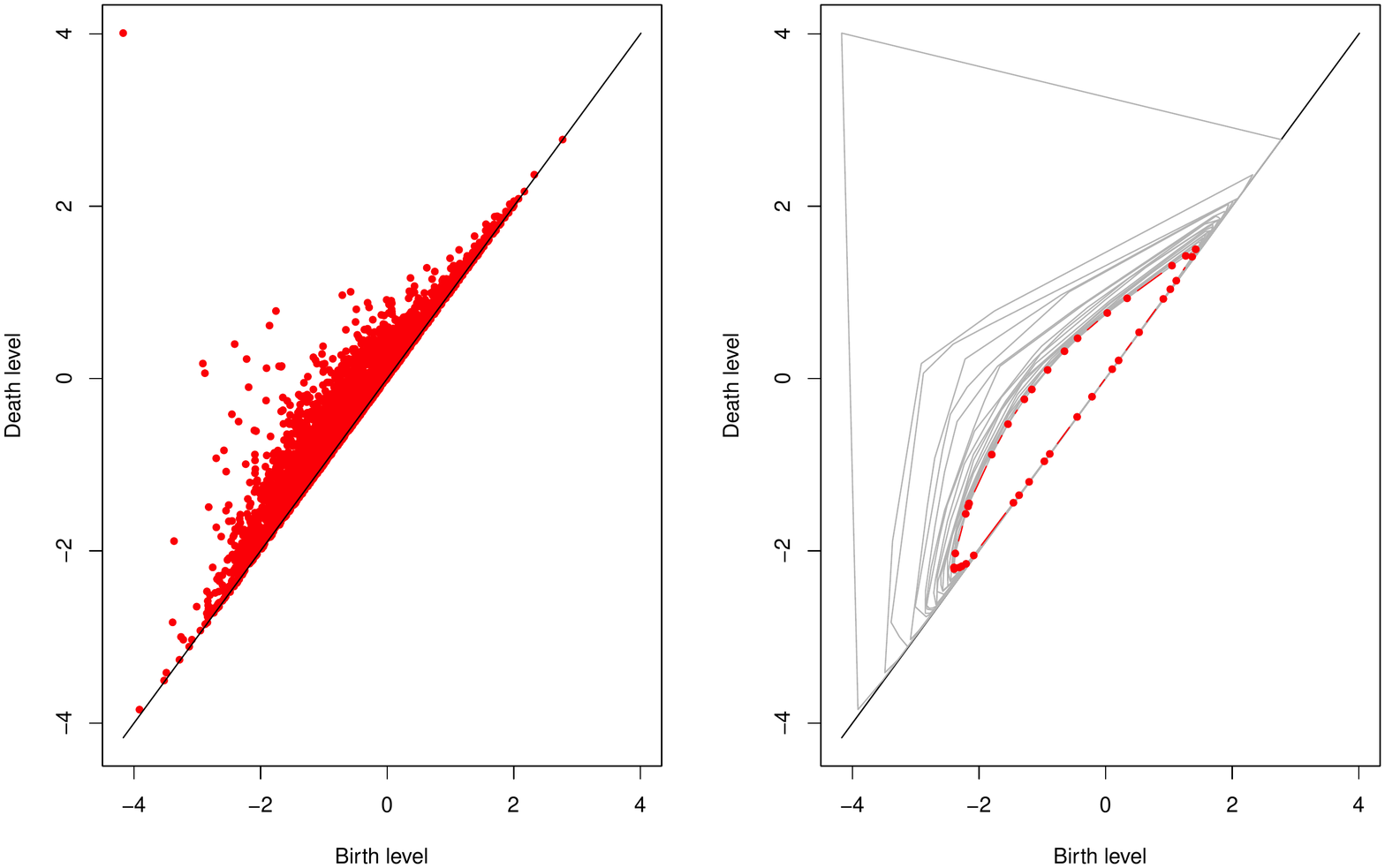}
\end{center}
\caption{{\it Component persistence diagram (left) and 90\% convex peel (right, as dots) for Region 1 in the GASS data. Grey lines in the right panel represent the peeled convex hulls.} \label{fig:pers}}
\end{figure}

A {\it persistence diagram} is a scatterplot of birth levels against death levels  for features of interest, in our case either components or holes.  The left plot of Figure
\ref{fig:pers} shows a persistence diagram for components in Region 1 of the transformed GASS data.  The first component to be born is by construction the last to die, producing the single point in the top left.   Otherwise, the points are clustered in a loose oval. Points near the diagonal represent less persistent structures mostly associated with noise, whereas more significant features are usually associated with points away from the diagonal.

\subsection{Convex peels and summary statistics}

It is sometimes difficult to interpret or compare persistence diagrams, either because of the large number of points or the bunching of many points along the diagonal.  We propose peeling successive convex hulls \citep{barnett76} until only a prescribed proportion of points remain, as illustrated in the right panel of Figure \ref{fig:pers}. In this way we extract the general shape of the persistence diagram without undue influence of either outliers or the mass of points near the boundary.  We summarise the shape of the final convex hull by the following five statistics:

\begin{enumerate}
\item the two centroid coordinates $(C_b,C_d)$;
\item the  perimeter $P$;
\item the area $A$;
\item the filamentarity, defined as \citep{bharad00,makarenko15}
\[ F = \frac{P^2-4\pi A}{P^2+4\pi A},\;\;\;\;\;\;\;0 \leq F \leq 1. \]
Thus defined, $F=0$ for a circle and $F=1$ for a line.
\end{enumerate}

\subsection{Bottleneck distance}

The bottleneck distance provides a measure in the space of persistence diagrams and can be used as a quantitative 
summary of the difference between two persistence diagrams, $A$ and $B$ say. A readable description, and further information, is given by  \cite{fasy14}.

First, the points in $A$ are matched to the points in $B$. This means that each point in $A$ is mapped  either to a unique point in $B$, or to its projection onto the diagonal line of birth-death equality.  The same is true for $B$.  Use of the diagonal is necessary because $A$ and $B$ can have different numbers of points. 

The {\it cost} of a mapping from  a point $(a_{ix},a_{iy})$ in $A$ to a point  $(b_{jx},b_{jy})$ in $B$ is the norm
\[ c_{ij} = \max\{ \mid a_{ix}-b_{jx} \mid, \; \mid a_{iy} - b_{jy} \mid \}. \]
The total cost  of a matching $P$ between $A$ and $B$ is $C(P)=\sum_{i,j}c_{ij}$.  The bottleneck distance is
then 
\[ w(A,B) = \min_P C(P), \]
where the minimum is taken over all possible selections of the linked pairs.
This, the bottleneck distance,  is essentially the smallest cost that could be incurred in mapping all points of $A$ into all points of $B$. Calculation is  numerically expensive, though efficient algorithms are available. For our application, we used the {\tt tda} package in R.

\section{Topology of a Gaussian random field on a lattice}
\label{sec:grf}

The persistent homology of a two-dimensional random field depends upon the number and distribution of local maxima and minima. In this section, we investigate the number of local extrema in a Gaussian random field on a  $d\times d$ lattice.  The results will be used in Section 6 to benchmark an analysis of the H\,{\sc i} data.  We assume stationarity, isotropy and N(0,1) margins throughout.

 The density of  critical points of a Gaussian random field in continuous space is a well-studied problem 
\citep{bard86}. It depends upon the joint Gaussian distribution of function values and first and second derivatives.  We are not aware of published  results for local extrema of a field on a discrete lattice however, although there has been work on the distribution of the global maximum over a region \citep{taylor07}.

\subsection{Expected number of local extrema}

Given a field $z(x)$, let $N_0$ and $N_1$ be the numbers of points in persistence diagrams for components and holes respectively.  
As previously stated, each point in a persistence diagram for components  corresponds to a local minimum of the field, and each point in a persistence diagram for holes corresponds to a local maximum. Due to  the symmetry of the Gaussian distribution, the number of local minima has the same statistical properties as the number of local maxima.  Because it is slightly tidier notationally, we will consider local maxima in this section.

Let $z_1=z(x_1)$ be the (scalar) field value at some location $x_1$ and let $z_2=z(x_2)$ be the $k$-dimensional vector of field values at the  immediate neighbours $x_2$ of $x_1$, with the neighbourhood to be defined later. Then, because we have standardised,  $(z_1,z_2)$ is Gaussian, with zero mean and  variance matrix
\[  \left(\begin{array}{cc}
 1 & r^T \\
  r & R
\end{array}
\right).\]
Here the $k$-vector of correlations $r$ between $z_1$ and $z_2$, and the $k \times k$ correlation matrix $R$ of $z_2$, each depend on the locations $x_1$ and $x_2$, though this is suppressed in the notation.

We have a local maximum at $x_1$ if $z_2 < z_1\one_k$ where $\one_k$ is a $k-$vector of ones. We can calculate this from the conditional distribution of $z_2$ given $z_1$ and the marginal distribution of $z_1$. First let $\phi^{(p)} (x; \mu, \Sigma)$ be the $p$-dimensional Gaussian probability density with mean vector $\mu$ and covariance matrix $\Sigma$, with $\Phi^{(p)} (x; \mu, \Sigma)$ being the corresponding cumulative distribution function. So $z_1$ has density  $\phi^{(1)} (z_1; 0, 1)$, $z_2$ has density
 $\phi^{(k)} (z_2; \zero_k, R)$, where $\zero_k$ is a $k-$vector of zeros, and the conditional density for $z_2$ given $z_1$ is
\[ \phi^{(k)} (z_2; rz_1, R-rr^T).\]

Then

\begin{eqnarray*}
p_1=P(\mbox{local maximum at } x_1) & = & P(z_2 < z_1\one_k) \\
\\
   & = & \int_{z_1} \left\{ \int_{z_2 < z_1\one_k}  \phi^{(k)} (z_2; rz_1, R-rr^T)\,dz_2 \right\}\phi^{(1)} (z_1; 0, 1)\,  dz_1 \\ 
\\
 & = & \int_{z_1}   \Phi^{(k)} (z_1\one_k; rz_1, R-rr^T)     \phi^{(1)} (z_1; 0, 1)\,  dz_1 \\ 
\\
 & = & \int_{z_1}   \Phi^{(k)} ((\one_k-r)z_1; \zero_k, R-rr^T)     \phi^{(1)} (z_1; 0, 1) \, dz_1. \\ 
\end{eqnarray*}

A general skew-Normal distribution described by 
\citet{gupta04, azzalini05, arnold09} and \cite{barrett14}
can be used to show that
\begin{equation}
\Phi^{(q)}(D\mu; \nu, \Delta+D\Sigma D^T)  = \int_y  \Phi^{(q)}(Dy; \nu, \Delta)  \phi^{(p)} (y ; \mu,\Sigma) \, dy,
\label{eqn:fx}
\end{equation}
where $\mu$ is of dimension $p$, $\nu$ is of dimension $q$, $\Sigma$ and $\Delta$ are $p \times p$ and $q \times q$ covariance matrices respectively, and $D$ is an arbitrary $q \times p$ matrix. Hence, if we set
\[  p=1,\;\;\;q=k,\;\;\;\mu=0,\;\;\;\Sigma=1, \;\;\;\nu=\zero_k,\;\;\; \Delta=R-rr^T,\;\;\;y=z_1 \mbox{ and } D=\one_k-r, \]
then
\begin{eqnarray}
 p_1 &  = &  \Phi^{(k)}(\zero_k; \zero_k,R-rr^T+(\one_k-r)(\one_k-r)^T) \nonumber \\
 & = &  \Phi^{(k)}(\zero_k; \zero_k, R+\one_k\one_k^T-\one_k r^T-r \one_k^T).
\label{eqn:expn}
\end{eqnarray}

Summing over all locations  yields $E[N_1]=E[N_0]=\sum_j p_j$, where $p_j$ is obtained from the equivalent of (\ref{eqn:expn}) for location $x_j$ instead of $x_1$. To calculate, we next need to define the neighbourhood of a lattice node.  Thinking of a lattice as an array of square pixels, those with common boundaries are clearly neighbours.  With this choice, an  interior pixel thus has $k=4$ neighbours, with differences in location of the form $(\pm 1, 0)$ or $(0, \pm 1)$.  We will call this the {\it cross} neighbourhood, which is what was used in the discussion of connected locations in Section 2.1.  We may alternatively  consider in addition pixels that share at least one vertex as neighbours. An interior pixel then has $k=8$ neighbours, consisting of the previous four and the four corners $(\pm 1, \pm 1)$. We will call this the {\it square} neighbourhood.   

Table \ref{tab:expn} shows $E[N_1]$ for three different grid sizes, assuming exponential correlation function with correlation length (the integral of the correlation function) equal to 20, which is close to that  in the H\,{\sc i} data.  We used the {\tt pmvnorm} routine in the {\tt mvtnorm} package within R, with the Miwa algorithm to calculate $\Phi^{(k)}$, and we adjusted neighbourhoods appropriately for pixels on boundaries.  The table also gives the means and associated standard errors of counts from 1000 simulations in each case.  The simulation results match the analytical results well. For larger  correlation lengths of $z(x)$ (not shown), there are fewer maxima or minima for a fixed size lattice, and the opposite for shorter correlation lengths. At zero correlation length, the number of minima (or maxima) is about 1/5 of the total number of points for the cross neighbourhood, and 1/9 for the square, as expected.

\begin{table}
\caption{{\it Expected number of local maxima for a discrete Gaussian random field on a $d \times d$ lattice, with exponential correlation function of correlation length 20. The simulation results are the  means (standard errors) from 1000 repetitions.} \label{tab:expn}}
\begin{center}
\begin{tabular}{lcccc}
              & \multicolumn{4}{c}{Neighbourhood} \\
              &\multicolumn{2}{c}{Cross, $k=4$} & \multicolumn{2}{c}{Square, $k=8$}\\
              & Simulation & Eq.(\ref{eqn:expn}) & Simulation & Eq.(\ref{eqn:expn})\\
$d=32$        &   128.3  & 128.4 & 68.8 & 68.9\\
              &  (0.3) & & (0.2)\\
$d=64$        & 498.3 & 497.9 & 258.3 & 258.3    \\         
              & (0.5)  & & (0.4)\\
$d=256$       & 7786.7 & 7786.5 & 3927.4 & 3929.5\\ 
              & (2.2) & & (1.6)\\
\end{tabular}
\end{center}   
\end{table}

\subsection{Variance of number of local extrema}

Let $I_j$ be an indicator of a local maximum at location $x_j$.  In order to determine the variance of $N_1=\sum I_j$ we need $E[I_iI_j]$ for all pairs of locations $x_i$ and $x_j$.  First we introduce some notation.
Let $z_1=(z_{11}(x_{11}), z_{12}(x_{12}))$ be the (bivariate) field value at any two locations
 $x_{11}$ and $x_{12}$. Let $z_{21}(x_{21})$ be the $k_1$-dimensional vector of field values over the neighbours $x_{21}$ of $x_{11}$, and let $z_{22}(x_{22})$ be the $k_2$-dimensional vector of field values over the neighbours $x_{22}$ of $x_{12}$.  Finally, let $z_2$ be the $(k_1+k_2)$-vector made up of  $z_{21}(x_{21})$ and $z_{22}(x_{22})$.

We need to consider separately the three possible arrangements of the neighbours:  first, when $x_{11}$ and $x_{12}$ have no common neighbours; second, when a neighbour is shared between $x_{11}$ and $x_{12}$;  and third,  when one of $x_{11}$ and $x_{12}$ is itself in the neighbourhood of the other.
In the first case, of no common neighbours,  $x_{21}$ and $x_{22}$ are distinct and $(z_1,z_2)$ is Gaussian with zero mean and  variance matrix
\[  \left(\begin{array}{cc}
 R_{11} & R_{12}^T \\
  R_{12} & R_{22}
\end{array}
\right),\]
where the sub-matrices once more depend on the locations, though this is still suppressed in the notation. 
Similarly to the previous section, we can use (\ref{eqn:fx})  to show that
\begin{equation}
 E[I_1 I_2]= \Phi^{(k)} (\zero_k; \zero_k; R_{22}-R_{12}  R_{11}^{-1}  R_{12}^T +DR_{11}D^T),
\label{eqn:i1i2}
\end{equation}
where $k=k_1+k_2$, $D= J -  R_{12}R_{11}^{-1}$ and $J$ is the $k\times 2$ matrix
\[ J  = \left(\begin{array}{cc}
 \one_{k_1} & \zero_{k_1} \\
\\
\zero_{k_2} & \one_{k_2}
\end{array}\right).   \]
From this we get the covariance between any pair of indicators $I_1$ and $I_2$ provided 
there is no neighbour in common. 

The second arrangement is when  $x_{21}$ and $x_{22}$ share one or more points so that $x_{11}$ and $x_{12}$  have at least one common neighbour. In that case $R_{22}$ is singular and we might anticipate difficulties. However, the variance matrix $R_{22}-R_{12}  R_{11}^{-1}  R_{12}^T +DR_{11}D^T$ in Equation (\ref{eqn:i1i2}) is {\it not} singular, at least in general, and  (\ref{eqn:i1i2}) still applies.

The final arrangement is simple. When either $x_{11} \in x_{22}$ or $x_{12} \in x_{21}$, then clearly there cannot be a local maximum at each of $x_{11}$ and $x_{12}$, so $E[I_1I_2]=0$ and Cov($I_1,I_2)=-E[I_1]E[I_2]$.

In principle, the variance of $N_1=\sum I_i$ can now be calculated.  However, the number of terms involved in the $d^2 \times d^2$ covariance matrices is unmanageable for lattices of large dimension $d$. Nonetheless, if the separation between pixels $x_i$ and $x_j$ is not small,  the covariance between $I_i$ and $I_j$ is negligible.    A working assumption of ignoring covariances between locations which are  separated by at least some threshold $\delta_0$ seems reasonable. Hence, our proposed estimator is
\begin{equation}
 \widehat{\mbox{Var}}(N_1)  = \sum_i \left\{E[I_i](1-E[I_i])+\sum_{j \in \calN_i}\mbox{Cov}(I_i, I_j) \right\},
\label{eqn:varhat}
\end{equation}
where $\calN_i=\left\{j: \mid x_i-x_j \mid \leq \delta_0\right\}$. In the following, we took $\delta_0=3$ lattice distance units.  The approximation is used together with Equation (\ref{eqn:i1i2}) in Table \ref{tab:sdn}. We used the GenzBretz algorithm  within the {\tt pmvnorm} routine for the covariances,  because the dimension of $R_{22}$ is too large for the Miwa routine.  The GenzBretz method involves Monte Carlo evaluations and hence leads to some uncertainty.   Nonetheless the agreement between the Monte Carlo results and our approximation is good.  The variance of $N_1$  is a little overestimated at $d=32$ and underestimated at $d=256$. More refined approximations with wider neighbourhoods and more careful treatment of boundary effects might give improvements, but the above seems adequate.  

\begin{table}
\caption{{\it Standard deviation of the number of local maxima of a discrete Gaussian random field on a $d \times d$ lattice, with an exponential correlation function of correlation length 20. Simulation results are standard deviations from 1000 simulations, while the approximation is based on equations (\ref{eqn:expn}), (\ref{eqn:i1i2})  and (\ref{eqn:varhat}). }\label{tab:sdn}}
\begin{center}
\begin{tabular}{lcccc}
              & \multicolumn{4}{c}{Neighbourhood} \\
              &\multicolumn{2}{c}{Cross, $k=4$} & \multicolumn{2}{c}{Square, $k=8$}\\
              & Simulation &  Approximation & Simulation &  Approximation\\
$d=32$        & 8.6  & 9.0 & 5.9 & 6.6 \\
      
\\
$d=64$        & 17.2  & 17.2  & 12.2 & 12.2   \\       
\\
$d=256$       & 70.7  & 66.0 & 49.1 & 45.2 \\

\end{tabular}
\end{center}  
\end{table}

\section{Topology of non-Gaussian random fields}
\label{sec:fields}

We are interested in assessing whether a topological approach can  be used to distinguish random fields when traditional methods fail.  For this purpose,  we explore properties of five different distributions on lattices, each stationary and isotropic, with N(0, 1) marginal distributions at the individual pixel level, and with correlation functions that are indistinguishable given  a single realisation. Thus, their quite different higher-order characteristics  would not be identified through analysis of first- and second-order properties.

The first model is a Gaussian random field (GRF) which we use as a reference. The other models are easily generated functions of one or more GRFs,  but are not themselves either GRFs or back-transformable to GRFs.
  We stress that the four non-Gaussian random fields are used for illustration only, and we do not claim any to be appropriate for the H\,{\sc i} data.  

\subsection{Simulated random fields}

The five distributions are constructed as follows.

\begin{description}
\item Model 1:  GRF $z$, with N(0,1) margins and exponential correlation function of correlation length $\eta$:
\[ \mbox{Corr}\left(z(x),z(x+d)\right)= \exp\left(-|d|/\eta\right). \]
Our default is  $\eta=20$.

\item Model 2:  $\chi^2_1$.  We start with a GRF $z_1$ with Matern correlation structure
\[ \mbox{Corr}\left(z_1(x),z_1(x+d)\right)=\frac{2^{1-\nu}}{\Gamma(\nu)}\left(\sqrt{2\nu}|d|/\eta\right)^\nu K_\nu\left(\sqrt{2\nu}|d|/\eta\right), \]
where $K_\nu(.)$ is a modified Bessel function of the third kind.  For  $\nu=0.5$ the Matern correlation function reduces to exponential. We construct a $\chi^2_1$ field as $z^\ast=z_1^2$ and then marginally transform using $z=\Phi^{-1}\left(F_{\chi^2_1}(z^\ast)\right)$, where $\Phi(.)$ and $F_{\chi^2_1}(.)$ are the N(0,1) and $\chi^2_1$ cumulative distribution functions, respectively.

\item Model 3:  $\chi^2_3$.  We begin with three independent GRFs $z_1, z_2, z_3$, each with the same Matern correlation structure.  We transform as
\[ z^\ast = \sum_{i=1}^{3} z_i^2, \hspace{1cm} z= \Phi^{-1}\left(F_{\chi^2_3}(z^\ast)\right). \]

\item Model 4: $T_3$.  We use four GRF with Matern correlation and our transformation is
\[ z^\ast= \frac{z_1}{\left(\sum_{i=2}^4 z_i^2/3\right)^{1/2}},   
\hspace{1cm} z= \Phi^{-1}\left(F_{T_3}(z^\ast)\right). \]

\item Model 5: $F_{3,3}$.  This time we have six GRF and take
\[ z^\ast= \frac{\sum_{i=1}^3 z_i^2/3}{\sum_{i=4}^6 z_i^2/3},  
\hspace{1cm} z= \Phi^{-1}\left(F_{F_{3,3}}(z^\ast)\right). \]
\end{description}

By construction, the marginal distributions are all N(0,1).  We estimated the Matern parameters $\nu$ and $\eta$ for Models 2--5 numerically, so as to match as closely as possible the computed correlation functions of the final fields $z$ to that of the reference Model 1.
Parameter estimates and simulation results for correlations at various distances are given in Table \ref{tab:cors}.  The mean results for Models 2--5 match Model 1 closely, well within sampling variation.  We contend that, as intended, these distributions cannot be distinguished through sample correlations on a single $256 \times 256$ lattice.

\begin{table}
\caption{{\it Parameter estimates and mean (standard deviation) correlations in the distance range 1--50 
for five types of random field.  Correlations results for Model 1 (Gauss) are exact, the others are empirical estimates based on 50 simulated $256 \times 256$ random fields with the given parameters.  }\label{tab:cors}}
\begin{center}
\begin{tabular}{lrrccccccc}
 & & & \multicolumn{7}{c}{Distance}\\
Model &   $\nu$ & $\eta$ &    1&     2&     3&     5&    10&    25&    50\\
1: Gauss      &0.50 & 20 &0.950& 0.905& 0.861& 0.779& 0.607& 0.287& 0.082\\
\\
2: $\chi^2_1$ &0.74& 41  &0.946& 0.894& 0.846& 0.755& 0.565& 0.220& 0.029\\
              &    &     &(0.012)& (0.022)& (0.031)& (0.046)& (0.076)& (0.106)& (0.083)\\
3: $\chi^2_3$ &0.54& 42  &0.952& 0.903& 0.857& 0.770& 0.590& 0.264& 0.033\\
              &     &    &(0.009)& (0.017)& (0.025)& (0.040)& (0.066)& (0.095)& (0.068)\\
4: $T_3$      &0.58 & 22 &0.950& 0.900& 0.851& 0.763& 0.585& 0.260& 0.049\\
              &     &    &(0.006)& (0.013& (0.018)& (0.028)& (0.047)& (0.076)& (0.082)\\
5: $F_{3,3}$  &0.54 & 50 &0.948& 0.897& 0.850& 0.762& 0.584& 0.272& 0.063\\
              &     &    &(0.009)& (0.017)& (0.024)& (0.037)& (0.061)& (0.080)& (0.061)\\
\end{tabular}
\end{center}
\end{table}

\subsection{Comparison of topological summaries}

\begin{figure}
\begin{center}
\includegraphics[width=7in,height=6in]{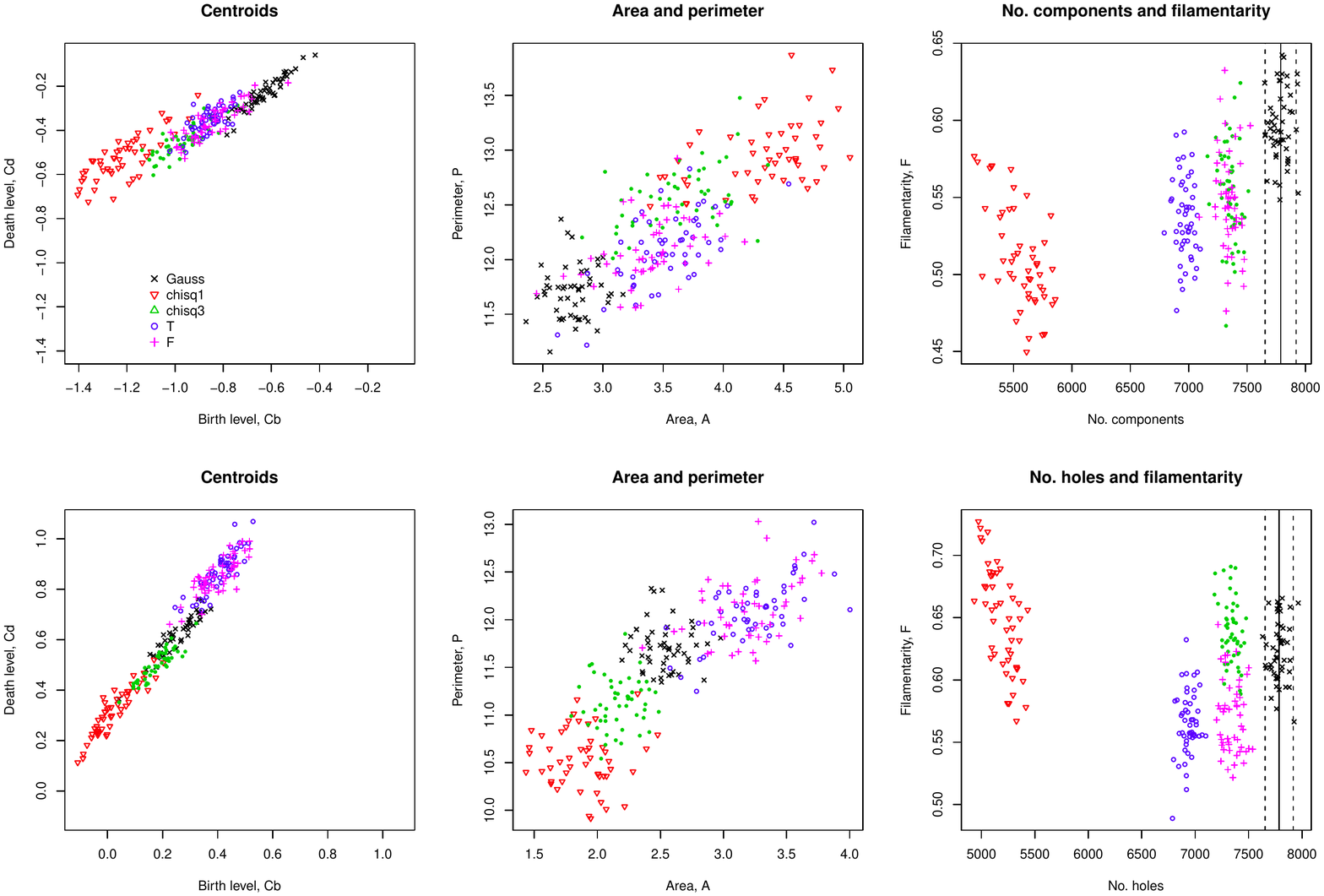}
\end{center}
\caption{{\it Number of structures and persistence diagram summaries for Models 1--5. Results from 50 realisations of $256\times 256$ fields, with the cross neighbourhood and 90\% convex peels.  The top row refers to components defined by local minima and the bottom row refers to holes  defined by local maxima. The solid vertical lines mark the expected count under a Gaussian random field, with the dashed lines indicating two standard deviations to each side.} \label{fig:tda}}
\end{figure}

Figure \ref{fig:tda} shows topological summaries of persistence diagrams for both components and holes,  for 50 simulated realisations of each model on  $256 \times 256$ lattices. We plot the total  number of points in the persistence diagram, either $N_0$ or $N_1$, together with the five simple statistics described in  Section \ref{sec:sums}, based on  90\% convex peels: the centroids, area, perimeter and filamentarity.  The top row is for the persistence diagrams defined by components, and the second row for holes. We used the cross neighbourhood: results for the square are similar.

The most obvious differences between the models is in the numbers of points, whether  $N_0$ or $N_1$. The solid vertical line in the right-most panels shows the expected value obtained in Section \ref{sec:grf}, with the broken lines marking two standard deviations on either side, again using the results of Section \ref{sec:grf}.  The theoretical values match the simulated data well and it is clear that of the models considered, the Gaussian random field has the largest numbers of both components and holes.  Model 2, based on $\chi^2_1$, has fewest structures of both types.    In most panels there is separation between at least some of the models, and there are some interesting contrasts between the rows.  For example, the $\chi^2_1$ model has the highest area and perimeter when components are considered, but lowest for holes. Also, there is evidence from the lower row  that Models 3 ($\chi^2_3$) and 5 ($F_{3,3}$) have similar counts of local maxima 
but differing values on the other summaries.  This suggests that there can certainly be information in the shape of a persistence diagram over and above the number of structures.

\section{Testing for differences between fields}
\label{sec:tests}

We can test an observed field against a Gaussian random field assumption by comparing the observed number of topological features with an interval based on the GRF properties. Figure \ref{fig:tda} suggests this may work well.
More generally, we might want to distinguish one field from another without assuming a particular distribution.  The previous section shows that differences in underlying structure might be detected through topological features but, without theoretical values or an underlying model, we need to develop a data-based approach.

Our proposal is to split the data from each field into subsets which are sufficiently separated for between--subset dependence to be weak, calculate appropriate summary statistics, and then use standard non-parametric statistical tests to compare the summaries from one field with another. Concentrating on $256\times256$ grids and correlation length around 20 as in the GASS data, we suggest splitting  each field  into nine $64\times64$ subsets with buffers of size 32  between them.  This gives a reasonable balance between the number of subsets (9) and correlation between them.  At the correlation length of 20 that we consider, correlation is 0.20 at the shortest distance between blocks (32), and 0.01 at the separation of block centres (96). Thus although the separate blocks are not independent, association is weak.

Our experience is that the most useful approach is to base 
tests  on two of the summaries:   counts $N_0$ of components and  filamentarity $F$ of persistence diagrams for holes. We suggest a 90\% convex peel  and use Wilcoxon tests to compare the 9 counts in one field with the 9 in another, and similarly the 9 filamentarities. A Bonferonni correction is used to adjust for the use of two tests.  The tests are of course not necessarily independent, and, as stated, we know the subsets are not independent.  Nonetheless the simulation results in Table \ref{tab:sims} suggest that performance is good. The standard deviation on test size is 0.01 and all estimated test sizes are within noise of the nominal 5\%.  It is clearly very easy to detect difference between $\chi^2_1$ fields and any of the others,  or between  Gauss and $T_3$.  There is good power for the other comparisons, except for $\chi^2_3$ against $F_{3,3}$, which are evidently hard to distinguish.  Power in this case can be increased by focussing if required only on filamentarity and not using the Bonfernonni correction.

\begin{table}
\caption{{\it Test size and power for nonparametric comparison between pairs of $256 \times 256$ random fields.  Each value is obtained from 500 simulated pairs, using 5\% overall tests after Bonferonni adjustment.  }\label{tab:sims}}
\begin{center}
\begin{tabular}{lrrrrr}
 & Gauss & $\chi^2_1$ & $\chi^2_3$ & $T_3$ & $F_{3,3}$\\
 
Gauss & 0.050  &1.000 & 0.782 & 1.000 & 0.776\\
$\chi^2_1$ &      & 0.062 & 1.000 & 1.000 & 1.000\\
$\chi^2_3$ &        &     &0.056 & 0.698 & 0.160\\
$T_3$ &             &     &      & 0.038  & 0.602\\
$F_{3,3}$   &       &     &      &        &  0.040\\
\end{tabular}
\end{center}
\end{table}

\section{Neutral Hydrogen Distribution in the Milky Way}
\label{sec:app}

We now return to the astronomical data, to understand whether the small-scale variation in the regions marked in Figure \ref{fig:gass} can be described by simple transfomations of a Gaussian random field, and whether the topologies of this variation can be considered to be the same in each region.

\begin{figure}[p]
\begin{center}
\includegraphics[width=\textwidth,height=5.5in]{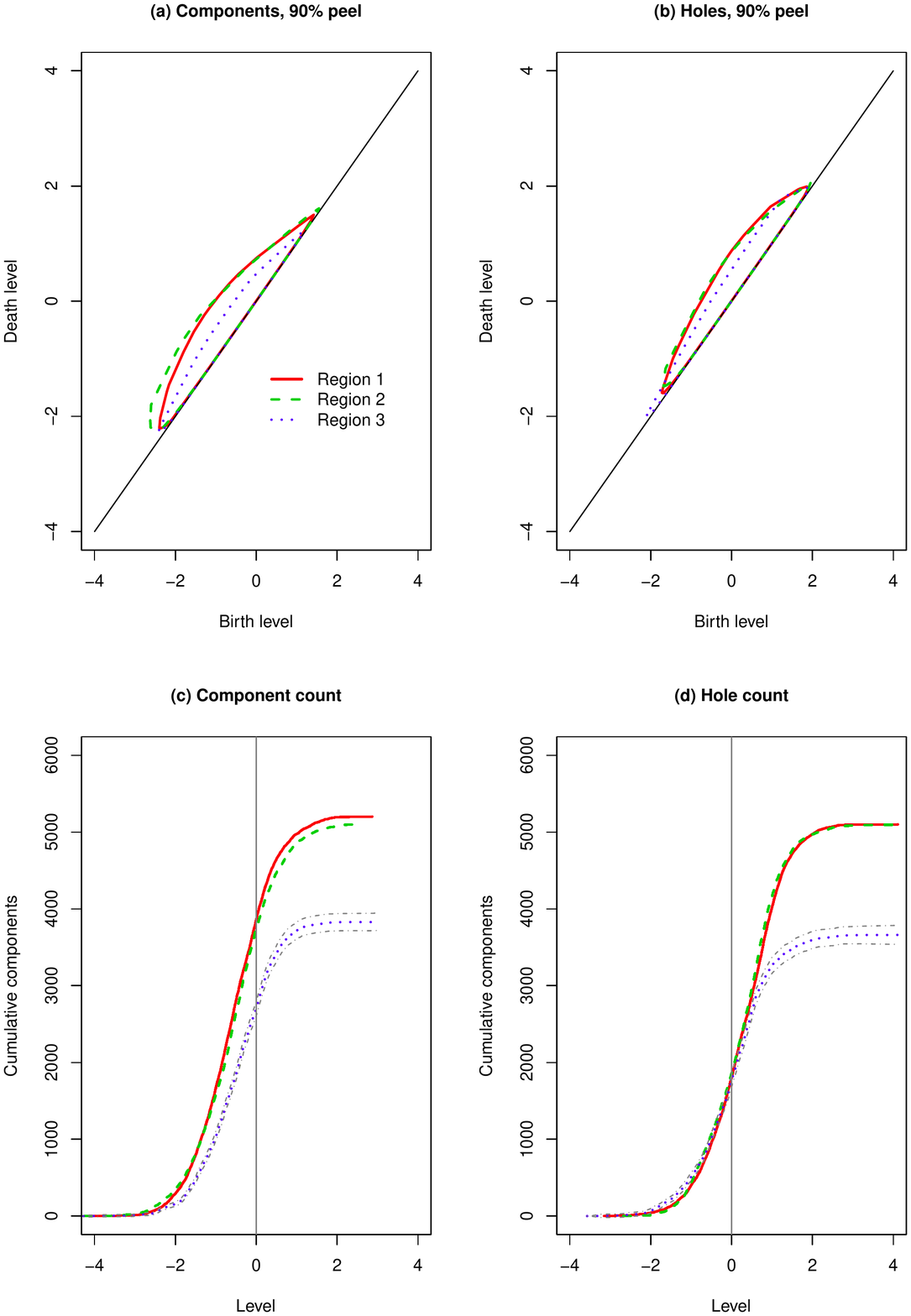}
\end{center}
\caption{{\it Topological summaries for the GASS data.  Regions 1, 2 and 3 as solid, dashed and dotted lines respectively.  The broken lines in the bottom row indicate two estimated standard deviations on either side of the Region 3 counts.  Standard deviations for the other regions are similar.} \label{fig:app}}
\end{figure}

Figure \ref{fig:app} shows, in the top row, 90\% convex peels of persistence diagrams for components and holes: it seems that in each case  the persistence diagram is more filamentary for Region 3 than the other regions. The bottom row of the figure shows cumulative counts of components and holes as the filtration $\calF_t$ increases. They include for Region 3 broken lines at plus and minus two estimated standard deviations around the count, obtained from the estimated predictable variation of 
the counting process \citep{and93}.  We have omitted similar intervals for Regions 1 and 2 as they confuse the plot, but we remark that their widths are very close to those for Region 3. Evidently, the cumulative counts are very similar for Regions 1 and 2 but very different for Region 3. This is confirmed by the final counts, which are given in Table \ref{tab:gasscounts}.  The table also gives the expected counts for a Gaussian random field with Matern correlation matching fits to the data.  Clearly, the observed counts are much lower than would be expected for a Gaussian random field, and the difference is stronger for Region 3.

Table \ref{tab:gasspvals} shows the $p$-values obtained from the data-splitting procedure of Section \ref{sec:tests}, based on filamentarity and counts.  Conclusions are as above, and our answers to both original questions are thus negative.  None of the regions has small-scale variation consistent with a simple transformation of a Gaussian random field, and Region 3 has markedly different structure from the other regions. 

\begin{table}
\caption{{\it Component and hole counts for the GASS data in the regions shown in Figure \ref{fig:gass}.  The first two numerical columns give estimated parameters for fitted  Matern correlation functions. Values in brackets are standard deviations obtained from estimated predictable variation.}\label{tab:gasscounts}}
\begin{center}
\begin{tabular}{lcccccc}
  & $\nu$  & $\eta$ & Components  & Holes & Gauss expected & Gauss SD\\
Region 1 &0.55 & 12.98 & 5201  & 5100  &   7896.1  & 65.9 \\  
         &     &       & (63.1)& (63.6)\\ 
Region 2 &0.70 & 13.50 & 5103  &  5095 &  7884.2 & 65.9 \\
         &     &       & (62.2)& (66.4)\\
Region 3 & 1.08& 14.90 & 3831  & 3661 & 7856.1 & 65.9\\
         &     &       & (57.9)& (60.5)\\
\end{tabular}
\end{center}
\end{table}

\begin{table}
\caption{{\it $P$-values for comparing the GASS Regions 1--3, based on data splitting into 9 sub-regions and Wilcoxon tests.}\label{tab:gasspvals}}
\begin{center}
\begin{tabular}{llcc}
              & Feature & Filamentarity & Number \\
Region 1 v Region 2 &  Components & 0.387 & 0.825\\
              &  Holes & 0.863 & 0.825\\
\\
Region 1 v Region 3 & Components & 0.077 & 0.040\\
              & Holes & 0.019 & 0.040\\
\\
Region 2 v Region 3 & Components & 0.222 & 0.027\\
              & Holes & 0.031 & 0.027\\
\end{tabular}
\end{center}
\end{table}

Finally for this section we note that the well-known bottleneck distance has no discriminatory power for these data.  Table \ref{tab:gassbotts} summarises bottleneck distances between all pairs of sub-regions, both within and between regions.  Mean intra-region distances can sometimes be larger than mean inter-region values, and any differences in means are small compared to the standard deviations.

\begin{table}
\caption{{\it Mean and standard deviation of bottleneck distances between persistence diagrams of sub-regions.  The column $M$ shows the number of pairs for each comparison.}\label{tab:gassbotts}}
\begin{center}
\begin{tabular}{llccccc}
      &    &     & \multicolumn{2}{c}{Components} & \multicolumn{2}{c}{Holes}\\
  \multicolumn{2}{c}{Comparison} & $M$ & Mean & SD & Mean & SD\\  
   Region 1 & Region 1&   36& 1.130 &0.574&    0.542& 0.144\\
   Region 1 & Region 2&   81& 1.237 &0.528&    0.671& 0.280\\
   Region 1 & Region 3&   81& 1.314 &0.558&    0.596& 0.131\\
   Region 2 & Region 2&   36& 1.486 &0.575&    0.799& 0.406\\
   Region 2 & Region 3&   81& 1.528 &0.620&    0.682& 0.365\\
   Region 3 & Region 3&   36& 1.287 &0.539&    0.533& 0.171\\
\end{tabular}
\end{center}
\end{table}

\section{Discussion}
\label{sec:disx}

Our analyses of the map of interstellar hydrogen (Fig.~\ref{fig:gass}), in terms of the shape of the persistence diagrams and the cumulative counts of topological features, identified Region 3 as being topologically distinct from the other two regions. The regions were deliberately selected without any prior knowledge of the properties of the interstellar medium in these parts of the sky. After carrying out the topological analysis we looked for an astronomical explanation for the distinctive nature of the gas in Region 3 and found that about half of this patch of the sky overlaps a nearby region of recent star formation known as the Orion-Eridanus superbubble \citep{naranan76, burrows93, pon16}. Stellar winds and/or supernova explosions have produced a bubble of hot gas, a few hundred light years across, whose boundaries can be traced in many different observational windows, including the H\,{\sc i} emission.  No structures due to a distinct astronomical object are present in the fields of Regions 1 and 2; in these parts of the sky, the distribution of the H\,{\sc i} is most likely the result of pervasive turbulent flows in the interstellar medium. Whilst recognising our identification of the Orion-Eridanus superbubble as retrospective, we speculate this as an explanation for the topological differences between Region 3 and either Regions 1 or 2.

The vast majority of analyses of the GASS or similar astronomical data rely on an assumption of an underlying  Gaussian random field, whether for the observational data with or without a simple transformation such as log, or as residuals around some large-scale structure. We have shown that all three of the regions of H\,{\sc i} that we examined contain strongly non-Gaussian fields, after trend removal. We suggest that statistical analysis of topological properties is an attractive alternative to existing approaches based on higher-order moments.  So far we have considered only two-dimensional data. Extension to three-dimensions may bring further power and is an important problem to address.

\bibliographystyle{agsm}
\bibliography{tdarefs}
\end{document}